\documentclass[prl,aps,showpacs,superscriptaddress,twocolumn, floatfix,
nofootinbib]{revtex4}

\usepackage{amsmath,amsfonts,amssymb}
\usepackage{graphicx}

\def\3{2.8in}    
\def\2{2.5in}
\def\4{3.0in}

\def \beq {\begin{equation}}
\def \eeq {\end{equation}}
\begin{document}

\title{Observation of Topological Order in the TlBiSe$_2$ class : Probing the "spin" and "phase" on topological insulator surfaces}
\author{S.-Y. Xu}\affiliation {Joseph Henry Laboratory of Physics, Department of Physics,
  Princeton University, Princeton, New Jersey 08544, USA}
\author{L.A. Wray}\affiliation{Joseph Henry Laboratory of Physics, Department of Physics,
  Princeton University, Princeton, New Jersey 08544, USA}\affiliation {Advanced Light Source, Lawrence Berkeley National Laboratory, Berkeley, California 94305, USA}
\author{Y. Xia}\affiliation {Joseph Henry Laboratory of Physics, Department of Physics, Princeton University, Princeton, New Jersey 08544, USA}
\author{R. Shankar}\affiliation {Joseph Henry Laboratory of Physics, Department of Physics, Princeton University, Princeton, New Jersey 08544, USA}
\author{S. Jia}\affiliation {Department of Chemistry, Princeton University, Princeton, New Jersey 08544, USA}
\author{A. Fedorov}\affiliation {Advanced Light Source, Lawrence Berkeley National Laboratory, Berkeley, California 94305, USA}
\author{J.H. Dil}\affiliation {Swiss Light Source, Paul Scherrer Institute, CH-5232, Villigen, Switzerland}\affiliation {Physik-Institute, Universitat Zurich-Irchel, CH-8057 Zurich, Switzerland}
\author{F. Meier}\affiliation {Swiss Light Source, Paul Scherrer Institute, CH-5232, Villigen, Switzerland}\affiliation {Physik-Institute, Universitat Zurich-Irchel, CH-8057 Zurich, Switzerland}
\author{B. Slomski}\affiliation {Swiss Light Source, Paul Scherrer Institute, CH-5232, Villigen, Switzerland}\affiliation {Physik-Institute, Universitat Zurich-Irchel, CH-8057 Zurich, Switzerland}
\author{J. Osterwalder}\affiliation {Physik-Institute, Universitat Zurich-Irchel, CH-8057 Zurich, Switzerland}
\author{R.J. Cava}\affiliation {Department of Chemistry, Princeton University, Princeton, New Jersey 08544, USA}
\author{M.Z. Hasan}\affiliation {Joseph Henry Laboratory of Physics, Department of Physics, Princeton University, Princeton, New Jersey 08544, USA} \affiliation{Princeton Center for Complex Materials, Princeton University, Princeton, NJ 08544, USA}\affiliation{Princeton Institute for Science and Technology of Materials, Princeton University, Princeton, New Jersey 08544, USA}

\pacs{}

\begin{abstract}

A topological insulator is characterized by spin-momentum locking on its boundary. The spin momentum locking on the surface of a three dimensional topological insulator leads to the existence of a non-trivial Berry's phase which leads to exotic transport phenomena on topological surfaces. Using spin-sensitive probes (Mott polarimetry), we observe the spin-momentum coupling and uncover the chiral nature of surface electrons in TlBiSe$_2$. We demonstrate that the surface electrons in TlBiSe$_2$ collectively carry a quantum Berry's phase of $\pi$ and a definite chirality ($\eta$ = -1, left-handed) associated with its spin-texture or vortex-structure on the Fermi surface on both the top and the bottom surfaces. Our experimental results for the first time not only prove the existence of Z$_2$ topological-order in the bulk but also reveal the existence of helical quasiparticle modes on the topological surface. Spin-texture calculations would be reported elsewhere.


\end{abstract}

\maketitle

\begin{figure*}
\includegraphics[width=18.5cm]{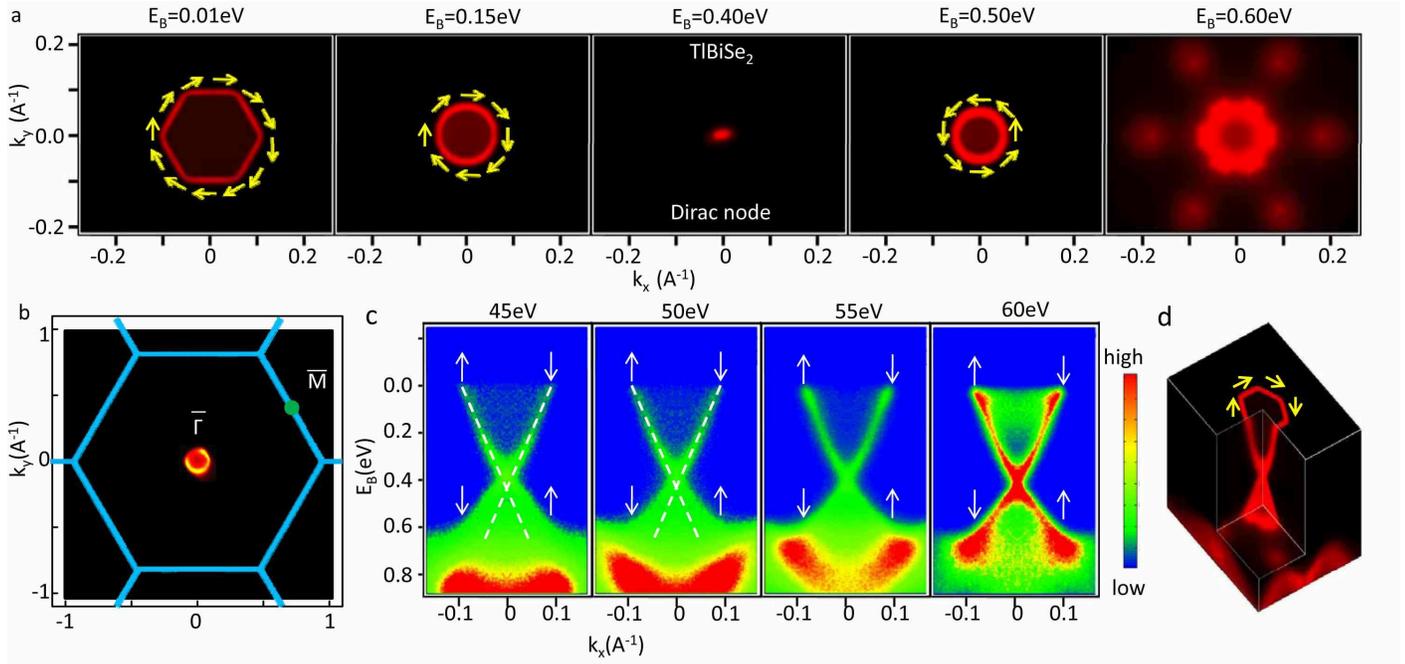}
\caption{\textbf{Observation of a spin-Helical topological ground state in TlBiSe$_2$.} a. Fermi surface topology evolution and spin vortex configuration through the Dirac point. Spins are drawn schematically (See Fig. 2 for spin polarization data) b. ARPES measurement over the First surface BZ on TlBiSe$_2$. All features are centered at the $\bar{\Gamma}$ point. Measurement is done in large step (low resolution) because a large momentum range needs to be covered. c. Incident-photo-energy dependence study of the surface band along $\bar{\Gamma}-\bar{M}$ direction, with spin directions on two branch of the Dirac bands. d. 3D representation of the dispersion of TlBiSe$_2$.}
\end{figure*}

Electron systems that possess a light-like conical Dirac spectrum, such as graphene \cite{Novoselov(Graphene), Checkelsky}, have recently been shown to harbor unusual collective states in high magnetic fields. Such states are possible because their light-like electrons come in spin-pairs that are chiral, which means that electrons moving in opposite directions have opposite pseudospins. An emerging direction in quantum materials research is the manipulation of atomic spin-orbit coupling to simulate the effect of a momentum dependent magnetic field, in attempt to realize novel spin phases of matter \cite{Day, Koralek, Moore Nature insight, Zahid RMP, Essin PRL Magnetic, ZhangDyon, Liang Fu PRL Superconductivity, Wray arXiv CuBiSe}. This effect has been proposed to realize systems consisting of spin polarized Dirac cones that are helical, meaning their direction of propagation is tied to the electron spin itself, which are forbidden to exist in ordinary Dirac materials such as graphene. Such spin polarized Dirac cones are predicted to exist at the edges of certain types of topological insulators and support exotic collective states that exhibit non-Maxwellian electromagnetic response \cite{ZhangDyon, Essin PRL Magnetic}, charge fractionalization and half-fermion states \cite{ZhangDyon, Liang Fu PRL Superconductivity, Wray arXiv CuBiSe} when interfaced with conventional electron systems. Because the surface electrons are localized to the very top layers, their effects are easily obscured by ordinary bulk-carriers. Three-dimensional topological insulators are a new phase of quantum matter that are realized by quantum entanglement effects \cite{Day, Moore Nature insight, Zahid RMP, Kane PRL, Moore PRB, David Nature BiSb, David Science BiSb, Matthew Nature physics BiSe,Zhang Nature physics BiSe, David Nature tunable, Franz Nature material Heusler, Zhang Nature material Heusler, Hsin Nature material Heusler, Phuan}, which were first experimentally observed in the Bi$_{1-x}$Sb$_x$ semiconductors \cite{David Nature BiSb}. In these systems, spin-orbit coupling gives rise to non-current carrying electronic states in the bulk and robust conducting states along the surfaces. In contrast to graphene, which has four Dirac cones (2 doubly degenerate cones at the K and K$'$ points in momentum space) \cite{Novoselov(Graphene),Checkelsky}, the remarkable property of topological surface states is that their dispersion is characterized by an odd number of spin-polarized Dirac cones. Odd spin surface metals are expected to exhibit a host of unconventional properties including a fractional (half-integer) quantum Hall effect and immunity to Anderson localization due to $\pi$ Berry's phase on their surfaces.


The key to testing these  experimental proposals is to find the most elementary form of helical spin-textured Dirac metals that are highly spin polarized \cite{David Science BiSb}. First principle band calculations suggest that the bulk of TlBiSe$_{2}$ belongs to the non-trivial topological class \cite{Eremeev JETP Thallium, Yan Euro Thallium, Hsin PRL Thallium}. Band-structure measured in photoemission qualitatively agrees with these band calculations \cite{Eremeev JETP Thallium, Yan Euro Thallium, Hsin PRL Thallium, Sato arXiv Thallium, Chen arXiv Thallium}. In this work, we investigate the existence of topological-order as defined via spin-momentum locking as Berry's phase via spin-resolved and electronic phase-sensitive investigation of single crystals of TlBiSe$_{2}$ for the first time. Such study is important for interpreting future spin-transport data from surface states which is expected to carry the hallmark of spin-texture. Remarkably, electrons on the surface of these materials are one-to-one spin-momentum locked with left-handed helicity carrying $\pi$ Berry's phase on the surface. The Fermi surface that is not mixed with any bulk states, suggests that these materials may be used for observing the effects of $\pi$ Berry's phase on the surface via transport measurements. As-grown samples of TlBiSe$_2$ exhibit residual metallic behavior in $\rho$(T) as in other topological materials. Therefore, it would be important to purify this material to reduce residual bulk contribution to conductivity. Very recently, it is becoming possible to isolate surface transport from that of the bulk in other topological insulators \cite{Phuan}. Similar methods may be applied here.

Spin-integrated angle resolved photoemission spectroscopy (ARPES) measurements were performed with $30 eV$ to $60 eV$ photon energy on beamline 12.01 and 10.01 at the Advance Light Source (ALS) in Lawrence Berkeley Laboratory. Spin-resolved ARPES measurements were performed at the SIS beamline at the Swiss Light Source (SLS) using the COPHEE spectrometer with two $40kV$ classical Mott detectors and the photon energy of $21eV$. Typical energy resolution was 10meV and about $1\%$ of the surface Brilliouin Zone (BZ) at Beamline 10 and 12 ALS and 80meV and $3\%$ of the surface BZ at SLS respectively. Samples were cleaved \textit{in situ} between $10$ to $80K$ at the Chamber pressure less than $5 {\times} 10^{-11} Torr$ at Beamline 10 and 12 at ALS and less than $2 {\times} 10^{-10} Torr$ at SLS respectively, resulting in shiny flat surfaces. Surface and bulk state band calculations were performed for comparison with the experimental data, using the LAPW method implemented in the WIEN2K package \cite{wien2k}.

The bulk crystal symmetry of TlBiSe$_2$ fixes a hexagonal surface Brillouin zone (BZ) for the cleaved (111) surface (Fig. 1b) on which $\bar{\Gamma}$ and $\bar{M}$ are the time reversal invariant momenta (TRIM) at which Dirac points can occur. Band structure measurements using ARPES are presented by scanning over the full BZ, in which all observed features occur in close proximity to the $\bar{\Gamma}$ point. High resolution dispersion maps traces a clear single Dirac cone for TlBiSe$_2$ (see Fig. 1d). The Dirac bands intersect the Fermi level at $0.116\AA$, with a particle velocity of $5.2{\times}10^5m/s$ along $\bar{\Gamma}-\bar{M}$ and cross $E_F$ at $0.1\AA$ along $\bar{\Gamma}-\bar{K}$ with a velocity of $4.7{\times}10^5m/s$. Due to the bend-bending effect \cite{David Nature tunable}, no band-like feature for bulk states can be clearly resolved inside the Dirac band. In order to systematically analyze the surface band structure of TlBiSe$_2$ and how it could be tuned with bulk doping, we perform a series of high resolution ARPES measurements of the constant energy contours at different binding energies (shown in Fig. 1a). The Fermi contour of TlBiSe$_2$ (constant energy contour at $E_B=0.01eV$) is hexagonally shaped, demonstrating the hexagonal warping effect \cite{Liang Fu Warping}. When the binding energy is increased from the Fermi level, the effect of the bulk potential vanishes and the shape of the contour reverts back to a circle. Lowering the binding energy further results in a Fermi surface of a single Dirac point with no other features. Constant energy contours below the Dirac point are observed to consist of the surface Dirac band, with an additional six-fold symmetric feature extending outside along all $\bar{\Gamma}-\bar{M}$ directions.

Modulating incident photon energy enables us to probe the $k_z$ dispersion of the electron kinetics, making it possible to distinguish surface from bulk photoemission features in a topological insulator \cite{David Nature BiSb, Matthew Nature physics BiSe}. We find that dispersion of the X-shaped Dirac band in TlBiSe$_2$ does not show any $k_z$ dependence (Fig. 1c), which verifies the surface origin of the Dirac band of a topological insulator. Variations in the quasi-particle intensity are observed, particularly inside the upper part of the Dirac cone and below $0.6eV$, indicating the nature of bulk conduction and valence band respectively.

Unlike conventional Dirac fermions as in graphene \cite{Novoselov(Graphene)}, helical Dirac fermions possess a net spin and are guaranteed to be conducting because of immunity to backscattering on the surface of topological insulators, allowing the unique possibility of carrying spin currents without heat dissipation in nanostructured materials. For a single Dirac cone topological insulator, the spin of the helical Dirac fermions fix the same direction on one branch of the X-shaped Dirac band, and take the opposite direction on the other branch (Fig. 1c). As a result, the spin vortex can be viewed as two dimensional counter-propagating spin system which follows the Fermi contour by certain chirality when looking at a constant energy plane. The observation of helical fermions characteries of quantum Berry's phase of $\pi$ carried collectively by the surface electrons, which distinguishes topological insulators from normal materials.

In order to measure the spin polarization, a spin-sensitive probe is required. Here we show that, using spin-resolved ARPES, we are able to study the helical electrons in TlBiSe$_2$ directly. Spin-resolved ARPES measurements \cite{Hoesch Spin ARPES} are performed using a double Mott detector (Au foil) configuration (Fig. 2a), which allows all three components of the spin vector of a photoelectron to be measured \cite{Hugo PRB}. Strong spin-orbit coupling in Au is known to create asymmetry in the process of photoelectrons scattering off Au foil, which will depend on the photoelectron's spin component normal to the scattering plane \cite{Hugo PRB}. Spin polarization in certain direction $P_{x,y,z}$ is then calculated by $P_{x,y,z}=(1/S_{eff}){\times}A_{x,y,z}$, where $S_{eff}$ is the Sherman function and $A_{x,y,z}$ is the asymmetry function defined from the asymmetry in the photoelectron scattering process.

\begin{figure}
\centering
\includegraphics[width=9cm]{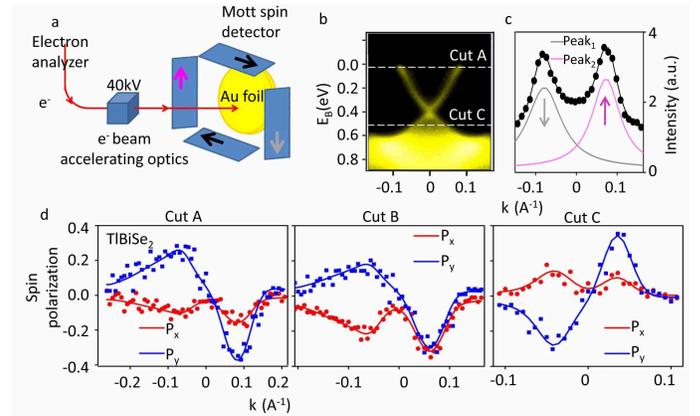}
\caption{\textbf{Chiral Spin-textures on the surface Dirac band topology of TlBiSe$_2$.}  a. Experimental geometry of the spin-resolved ARPES setup. The photoemission part is ignored in this schematic. Full spin directions are measured by two Mott spin detectors; only one is drawn here. b. ARPES spectrum along $\bar{\Gamma}-\bar{M}$ direction, showing the energy position of cut A, B and C. The directions of the cuts in momentum space are not necessarily along $\bar{\Gamma}-\bar{M}$ direction. c. Momentum-Distribution curve of spin-average spectrum of cut A at E$_B$=0.02eV, together with Voigt function peaks of the fit. d. Measured in-plane spin-polarization of cut A, B and C. The local coordinate system of each cut is different: for each cut, cut direction is defined as the x-axis, and the z axis is perpendicular to the sample plane.}
\end{figure}

We show three spin-resolved measurements at two different binding energy levels (Fig. 2b), which allow us to gain comprehensive understanding on the spin texture of TlBiSe$_2$. Spin-resolved measurements (cuts) are made along certain direction in 2-D momentum space on a constant energy plane. Cuts A and B are made at the experimental Fermi level, whereas cut C is on the circular constant energy contour range, below the Dirac point. We start by showing the in-plane spin polarization spectra ($P_x$ and $P_y$) of the three cuts (Fig. 2d). Peaks of the spectra correspond to the position where the cuts cross the constant energy contours. A local coordinate system is defined for each cut: the x axis is along the cut direction and the z axis is perpendicular to the sample plane. Clear polarizations in both x and y directions are observed (up to $40\%$). For each single cut (either A, B, or C), the y-direction component of
polarization has an opposite sign at opposite sides of the Fermi contour, revealing the helical Dirac fermion nature. Clear polarizations in x direction are also shown, though, of the same sign within each single cut. Considering $P_x$ (same sign) and $P_y$ (opposite sign) together, we are able to obtain detailed spin direction information: directions of the spins are perpendicular to the corresponding momentum vectors with left-handed chirality. Therefore, the surface of TlBiSe$_2$ carries a non-zero ($\pi$) quantum Berry's phase via the topological surface electrons.


Our spin and momentum-resolved spectroscopic studies on TlBiSe$_2$ enable us to probe the helical Dirac fermion on its surface. Clear spin polarization provides comprehensive information (e.g. direction and magnitude) of the spin vortex, by which we clearly demonstrate the quantum Berry's phase ($\pi$) carried by the surface electrons. The remarkable observation of helical Dirac fermions, which possess opposite spin when moving in opposite directions, serves as a critical evidence of immunity to backscattering and localization on the surface due to the time-reversal invariant nature of the $Z_2=-1$ topologically ordered material.

\newpage

\end{document}